# Cloaking of Arbitrarily Shaped Large-Scale Objects Through the Injection of Electromagnetic Invisibility Genes

Zirui Xie[1,#], Fei Sun[1,#,*], Yichao Liu[1,*], Jiale Li[1], Jianpu Yang[1], and Shuai Zhang[2]

*1 Key Lab of Advanced Transducers and Intelligent Control System, Ministry of Education and Shanxi Province, College of Physics and Optoelectronics, Taiyuan University of Technology, Taiyuan, 030024 China*

*2 The Department of Electronic Systems, Aalborg University, 9220 Aalborg, Denmark*

[#] Zirui Xie and Fei Sun contributed equally to this work.

[*] Corresponding author: sunfei@tyut.edu.cn liuyichao@tyut.edu.cn

## Abstract

Full-space electromagnetic invisibility can eliminate the scattering cross-section of an object to achieve perfect invisiblity, and its physical mechanisms mainly include two categories: light-bending cloaking and scattering-cancellation cloaking. Light-bending cloaking guides detecting waves around the target but disables the hidden object from perceiving external electromagnetic signals, leading to double-blind invisibility that is incompatible with sensing applications. Scattering-cancellation cloaking enables the hidden object to interact with external signals while achieving invisibility, making it highly suitable for electromagnetic sensors and communication systems. However, traditional scattering-cancellation cloaking relies on adding external structures to cancel scattering, which exhibit strong dependence on the shape and size of the target. It is difficult to realize cloaking for irregular, inhomogeneous and electrically large objects, and the structure fails once the shape, size, or material of the object changes, requiring complete redesign. To address these critical limitations, this work proposes an electromagnetic invisibility gene injection strategy motivated by biological cellular camouflage mechanisms. Arbitrarily shaped, sized, and composed objects are decomposed into subwavelength units, and a customized invisibility gene is designed and injected into each unit according to its electromagnetic properties. After reassembly, the entire object achieves efficient scattering-cancellation cloaking effect. Numerical simulations demonstrate that this method enables effective cloaking for objects with arbitrary macroscopic shapes, diverse dielectric constants ranging from 2 to 10, and different subwavelength unit boundary morphologies. Microwave-band experimental results further validate the feasibility and effectiveness of the proposed strategy, showing significant scattering suppression and ideal electromagnetic invisibility. This method breaks the limitations of traditional cloaking design and provides a universal, flexible and low-complexity solution for practical cloaking in antenna support structures, communication system mounts, and electromagnetic transparent protective covers.

## Introduction

In contrast to traditional stealth technology, which primarily absorbs or deflects radar waves in directions undetectable by radar, invisibility cloaking can simultaneously reduce both the scattering and absorption cross-sections of an object [1-5]. In recent years, the field of invisibility cloaking has expanded its scope from electromagnetic waves to include static magnetic fields [6, 7], heat flux [8, 9], water waves [10], acoustic waves [11], and even simultaneous invisibility across multiple physical fields [12-17]. Its application significance has also expanded from the initial cloaking of electromagnetic targets to encompass protection of sensitive biomedical components [6] flood prevention and control [10], and multi-physical field compatible management in chip systems [13], among other diverse fields [14].

Cloaking can be classified into full-space invisibility cloaking and half-space (carpet) cloaking. Half-space cloaking only disguises curved surfaces as flat planes, which does not constitute real invisibility. In contrast, full-space cloaking renders objects in free space genuinely invisible by reducing their scattering cross-section to zero, and thus represents the focus of this work. Full-space cloaking is generally realized through two mechanisms [2]: light-bending cloaking [1, 18-26] and scattering-cancellation cloaking [27-34]. Light-bending cloaking routes waves around the target by using specially engineered electromagnetic media, such as optic-null medium [19-26], and imposes almost no restriction on the shape, size, or material of the concealed object. However, this scheme isolates the object from external electromagnetic signals, leading to double-blind invisibility that is unsuitable for sensors and detectors [35]. Scattering-cancellation cloaking instead uses additional media to destructively interfere with the scattering of the hidden object, realizing zero total scattering while maintaining signal interaction with the target.

For scattering-cancellation cloaking, the concealed object can interact with the external environment (i.e., receiving/transmitting electromagnetic signals), thus achieving a ‘seeing without being seen’ effect that is well-suited for cloaking sensor/detector probes [35-38] and radiating antennas [39, 40]. However, such cloaks typically require full redesign when the shape or material of the target changes. Designing cloaks for complex or electrically large objects is especially difficult: Mie-theory-based schemes are limited to simple geometries such as spheres and cylinders [30], while optimization-driven approaches often produce cloak structures much larger than the target itself. As the shape complexity and electrical size increase, the computational cost and structural complexity rise sharply, making practical design infeasible. Consequently, despite its capability to enable perception-compatible invisibility, existing scattering-cancellation cloaking still faces critical bottlenecks: it is difficult to implement for irregular, inhomogeneous, and electrically large objects, and the cloak fails and requires complete redesign once the shape, size, or material composition of the target is altered

To address the critical challenge of designing scattering-cancellation cloaks for large-scale objects with arbitrary geometries—where conventional methods face redesign burdens for shape/material changes and computational complexity for irregular, inhomogensous, large targets—this study proposes an electromagnetic "invisibility gene" injection method. This approach enables universal scattering cancellation for objects of arbitrary size, shape, and material composition ($\varepsilon_o$ = 2–10), while preserving their ability to interact with external electromagnetic fields. Motivated by cellular-level adaptive camouflage mechanisms in organisms, we decompose irregularly shaped and large-scale target object into subwavelength-scale units. Tailored electromagnetic "genes" are then designed and injected into each unit according to its electromagnetic properties, collectively achieving high-performance scattering-cancellation cloaking effect for the entire structure (zero scattering cross-section). By decomposing complex targets into subwavelength units and implanting customized scattering cancellation genes, this strategy overcomes critical challenge of designing scattering-cancellation cloaks for large-scale objects with arbitrary geometries. It therefore enables practical, reliable cloaking for rigid supporting structures in antenna and communication systems, as well as electromagnetic transparent protective covers, without interfering with the normal signal transmission and reception functions of the protected devices.

## Results

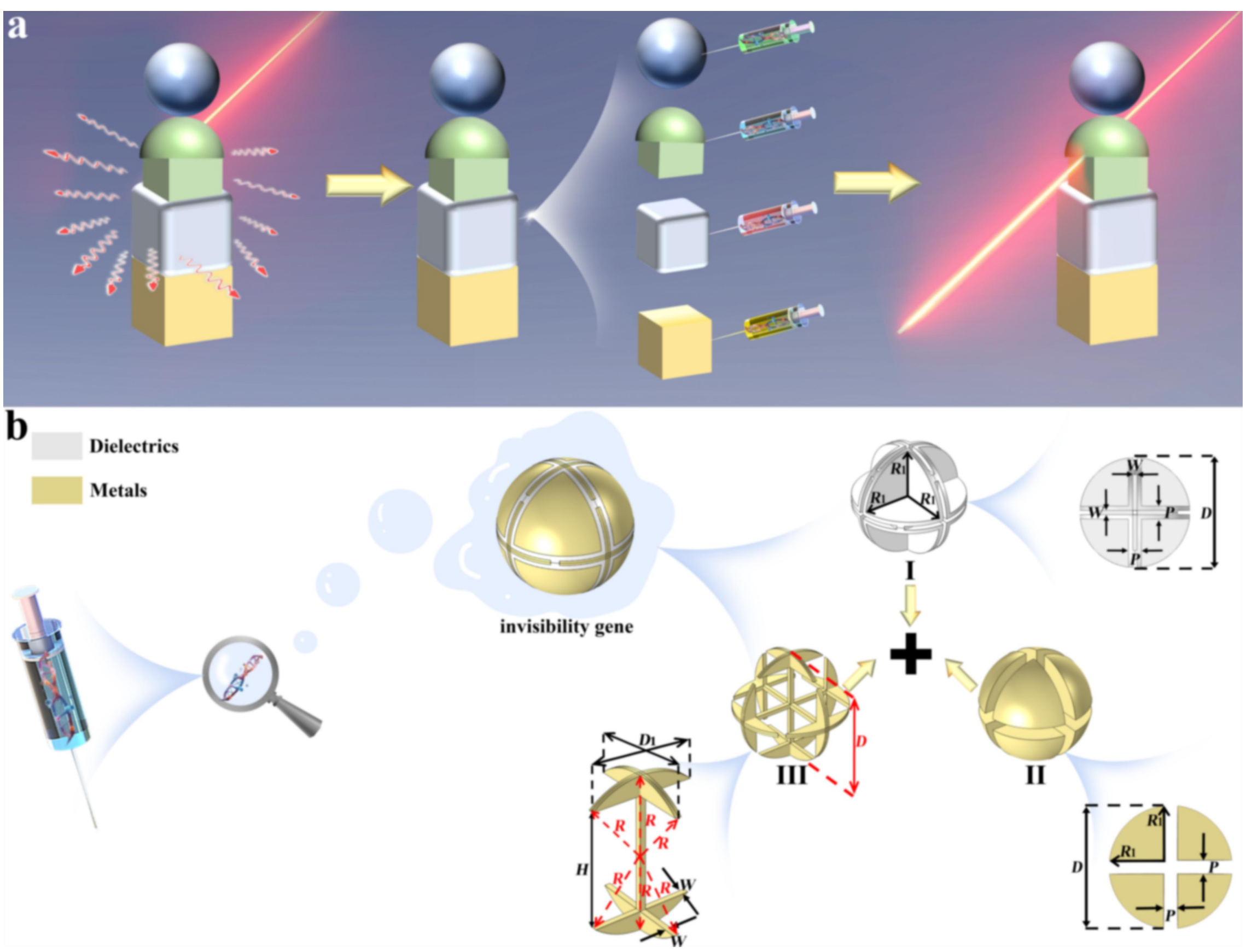

**Fig. 1| a** Flowchart for achieving invisibility of an irregularly-shaped, scattering-prone object composed of different materials by injecting the corresponding invisibility genes: Step 1: Divide the object to be cloaked into multiple subwavelength units based on the operating wavelength $\lambda_0$. Step 2: Design distinct invisibility genes according to each unit's material properties. Step 3: Inject the designed invisibility genes into corresponding subwavelength units, then reassemble these gene-injected units into the original shape of the object, thereby achieving cloaking invisibility effect of the entire object at the operating wavelength $\lambda_0$ (see **Supplementary Movie 1**). **b** Schematic of the three structural components (Structure-I, Structure-II, Structure-III) comprising the spherical invisibility gene injected into each subwavelength unit.

### Design Method for achieving invisibility by injecting invisibility genes

**Fig. 1a** illustrates the proposed invisibility gene injection methodology for achieving cloaking in arbitrarily shaped, large-scale objects. The schematic depicts a multi-material object (represented as a tower structure with colored blocks indicating different materials) exhibiting significant scattering (denoted by red wavy lines) under incident electromagnetic waves (red ray). To eliminate scattering, the object is decomposed into various subwavelength units (e.g., cubes, spheres, mushroom shapes)

—each with fixed isotropic permittivity (non-magnetic assumption) and each with fixed isotropic permittivity that may differ between units. Tailored invisibility genes are injected into these units, which are then reassembled into the original geometry. This process achieves near-zero scattering cross-section (**Supplementary Movie 1**), imposing no inherent constraints on object size, shape, or material composition—unlike traditional scattering-cancellation cloaking based on either Mie scattering theory or optimization algorithms. The approach enables scattering-cancellation cloaking in the microwave regime for arbitrarily shaped, large-scale non-magnetic dielectric objects ($\varepsilon_o$ = 2–10), preserves their electromagnetic signal reception capability, and facilitates potential cloaking applications for complex-shaped sensor probes.

The proposed invisibility gene injection method fundamentally operates on scattering-cancellation principles [27-39]. When illuminated by electromagnetic waves, a non-magnetic dielectric subwavelength unit (e.g., cubes, spheres, mushroom shaped units in the second step of Fig. 1a) undergoes polarization, exhibiting an induced electric dipole moment macroscopically. The core function of the invisibility gene is to precisely cancel this dipole moment. To achieve this, we embed an umbrella-shaped metallic scaffold (Structure-III in Fig. 1b) within each subwavelength dielectric unit. Under incident waves, plasmonic resonance in this scaffold generates a tailored electric dipole moment along its longitudinal axis. For omnidirectional cancellation, three such scaffolds are orthogonally interconnected, ensuring isotropic response in 3D space (Structure-III in Fig. 1b). Critically, Structure-III concurrently excites undesired transverse magnetic responses. To suppress these, we integrate eight diamagnetic metallic spheres (Structure-II in Fig. 1b), which induce counter magnetic moments to neutralize parasitic resonances. Furthermore, dipole moments induced in subwavelength units scale with their permittivity $\varepsilon_o$. We therefore encapsulate Structure-III with a dielectric shell (Structure-I in Fig. 1b) of engineered permittivity $\varepsilon_I$. This shell acts as a permittivity-compensating layer, dynamically offsetting permittivity-dependent dipole variations across units. The assembled invisibility gene (Structure-I+II+III in Fig. 1b) is injected into each subwavelength unit. At the target wavelength $\lambda_0$, this integrated system reduces the unit's scattering cross-section to near-zero (**Supplementary Movie 1**), achieving effective electromagnetic transparency.

In this study, for the standard cubical subwavelength unit with a side length of $L = 0.15\lambda_0$ obtained by decomposing the target object to be cloaked, a design scheme for the spherical invisibility gene with fixed geometric parameters and only dielectric parameter regulation is adopted. Among them, Structure-III serves as the core umbrella-shaped metallic scaffold structure, which consists of three identical scaffolds arranged orthogonally and concentrically along the $x$, $y$, and $z$ axes. The metallic rods have a length of $H$ and a square cross-section with a side length of $W$; the umbrella heads have a uniform thickness of $W$ and a unilateral length of $D_1$, providing a structural foundation for three-dimensional isotropic scattering cancellation. Structure-II is the parasitic magnetic response suppression structure, composed of eight identical eighth-spheres fabricated from diamagnetic metal, with a single eighth-sphere having a radius of $R_1$ and a spacing of $P$ between adjacent eighth-spheres, which precisely neutralizes the undesired transverse magnetic response of Structure-III. Structure-I is the permittivity compensation layer that covers around Structure-III. It is a non-magnetic dielectric shell with only the permittivity $\varepsilon_I$ regulated, acting as the core regulatory structure for adapting to units with different permittivities $\varepsilon_o$. Structure-II and Structure-III are set as perfect electric conductors in the microwave band simulations, and the permittivity $\varepsilon_I$ of Structure-I is determined by the permittivity $\varepsilon_o$ of the host unit (see Fig. 2). This standardized design not only ensures the scattering cancellation performance but also provides a fixed parameter basis for subsequent engineering fabrication.

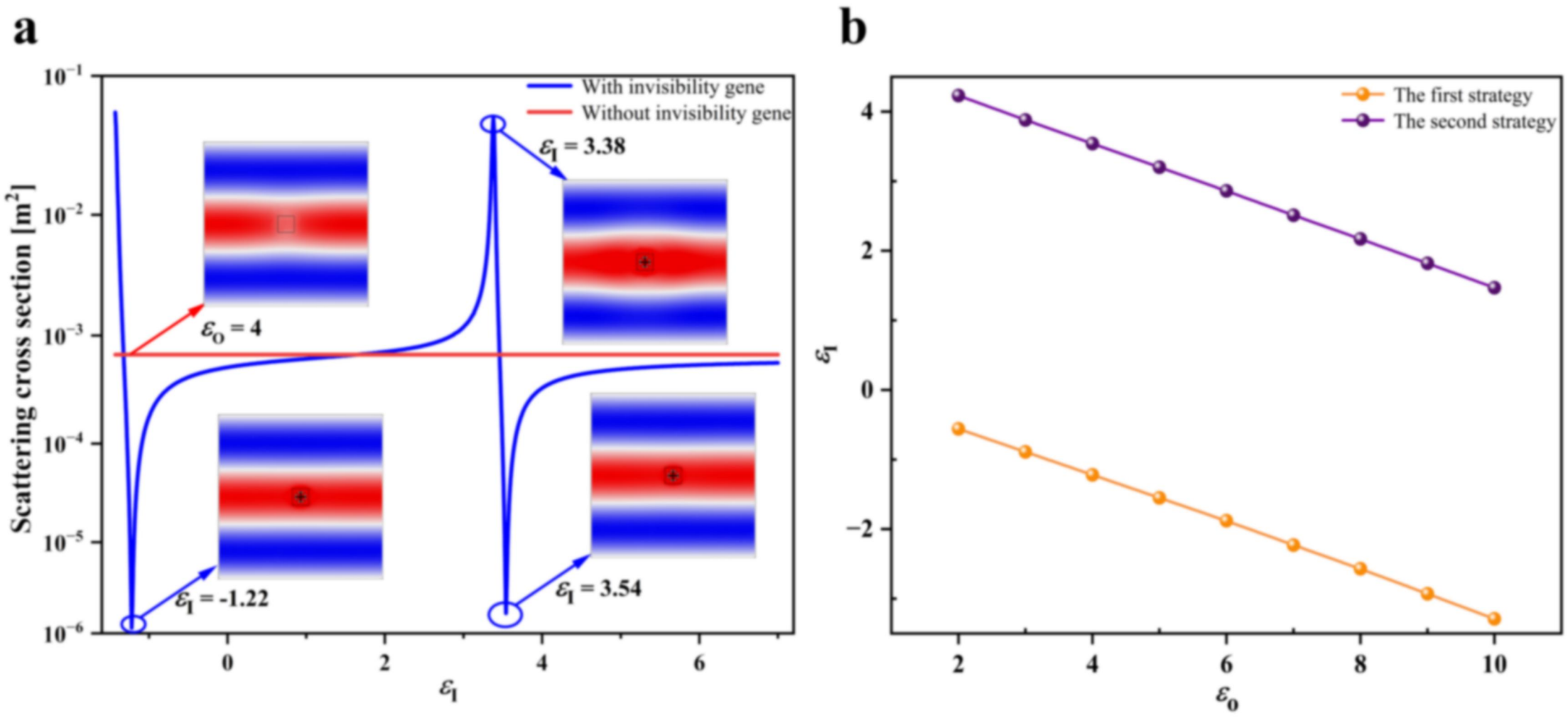

**Fig. 2| a** The correlation law between the permittivity $\varepsilon_I$ of Structure-I in the spherical invisibility gene and the scattering cross-section of the subwavelength unit after invisibility gene injection, obtained via numerical simulation. Simulation parameter settings: the permittivity of the cubical subwavelength unit to be cloaked is fixed at $\varepsilon_o = 4$, and the operating frequency of the incident electromagnetic wave is $f_0 = 1.7$ GHz. **b** Two sets of corresponding relationships between the intrinsic permittivity $\varepsilon_o$ of the cubical subwavelength unit and the optimal permittivity $\varepsilon_I$ of Structure-I in the spherical invisibility gene, when the scattering cross-section of the unit reaches the minimum after the injection of the spherical invisibility gene.

To verify the regulatory effect of the permittivity compensation layer (Structure-I) and clarify the influence law of $\varepsilon_I$ on the scattering cancellation performance of the subwavelength unit, numerical simulations are carried out in this study on a cubical subwavelength unit with an operating frequency of $f_0 = 1.7$ GHz ($\lambda_0 = 0.176$ m) and $\varepsilon_o = 4$, and the correlation law between $\varepsilon_I$ and the total scattering cross-section of the unit presented in Fig. 2a is obtained. When $\varepsilon_I$ takes the two characteristic values of -1.22 and 3.54 respectively, the scattering cross-section of the unit has a reduction rate close to 100%, achieving near-perfect scattering cancellation. On this basis, we further investigated the optimal $\varepsilon_I$ that minimizes the scattering cross-section of non-magnetic dielectric units with $\varepsilon_o$ in the range of 2 – 10, and obtained the dual-value mapping relationship between $\varepsilon_o$ and the optimal $\varepsilon_I$ shown in Fig. 2b (this dual-value mapping phenomenon is physically homologous to the dual-frequency resonance characteristics of multi-layer plasmonic cloaks [28]). The negative $\varepsilon_I$ can be realized by metal-based metamaterials in the microwave band, while the positive $\varepsilon_I$ can be realized by conventional dielectrics such as high-permittivity ceramics. This mapping relationship provides a core basis for the customized design of units with different dielectric properties. Meanwhile, the shape adaptability and parameter robustness of the spherical invisibility gene are verified: for irregular subwavelength units with dimensions close to the standard cubical unit, effective scattering cancellation can be achieved only by adjusting $\varepsilon_I$; in the interval near the optimal $\varepsilon_I$, the scattering cross-section of the unit can still be reduced by more than 80%, which provides sufficient fault tolerance for parameter deviations in actual fabrication.

The above research has clarified the scattering cancellation mechanism of the invisibility gene through principle design, determined the specific geometric parameters and material configuration of the three core structures through standardized structural design, verified the high-efficiency scattering cancellation effect of a single subwavelength unit injected with the customized invisibility gene via the numerical simulation in Fig. 2, clarified the key mapping law between $\varepsilon_o$ and $\varepsilon_I$, and completed the theoretical and simulation verification of unit-level invisibility. To further break through the limitations of traditional scattering-cancellation cloaking on the shape and size of the target and verify the universality of the proposed method for macroscopic targets, the full-wave numerical simulation shown in Fig. 3 is performed as follows: the macroscopic target to be cloaked with an arbitrary shape is decomposed into a number of subwavelength units, each unit is injected with the invisibility gene matching its material properties and then reconstructed into the original geometry. On this basis, we systematically verify the overall scattering cancellation effect of this method on dielectric targets with arbitrary shapes, and complete the feasibility verification of invisibility from the unit level to the macroscopic target level.

**Simulated Results**

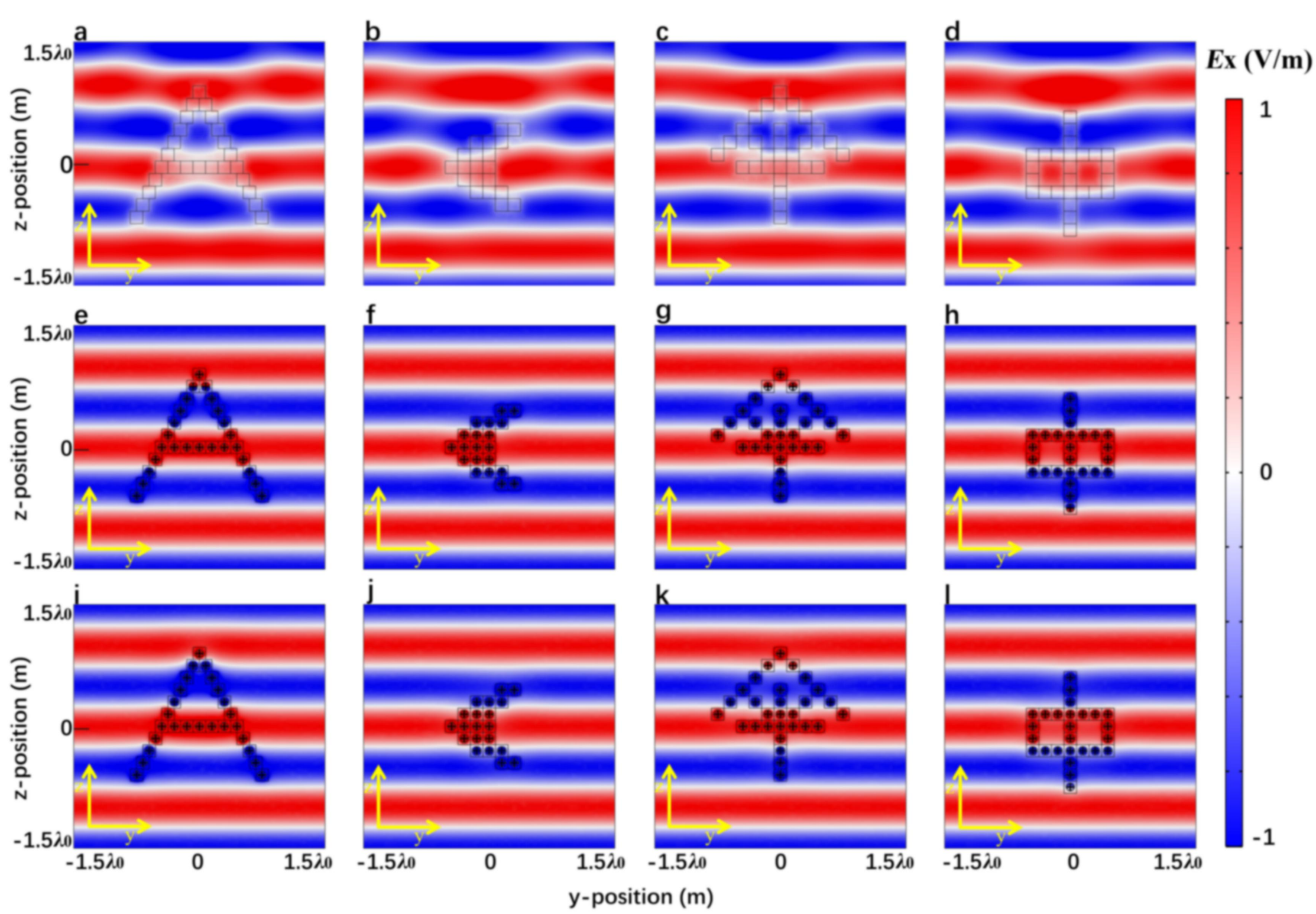

**Fig. 3| 3D simulated results: normalized simulated electric field's $x$-component $E_x$ on the $x$ = 0 cross-section when a TE-polarized plane electromagnetic wave of the frequency $f_0$ = 1.7 GHz is incident along the positive $z$-axis onto different shaped structures. a-d**, Structures shaped like character "A", crescent, umbrella, and chinese character "中", respectively, made of dielectric with permittivity $\varepsilon_o$ = 6 without injected invisibility genes. **e-h**, Corresponding structures to **a-d** after injecting invisibility genes using the first strategy (Structure-I with negative permittivity in Fig. 2b). **i-l**, Corresponding structures to **a-d** after injecting invisibility genes using the second strategy

(Structure-I with positive permittivity in Fig. 2b). In simulations, boundary conditions, geometric parameters, and material settings for the original objects and invisibility genes are detailed in the **Numerical Method** .

To verify that arbitrarily shaped objects can be rendered invisible by injecting the invisibility gene with the procedure illustrated in Fig. 1, Fig. 3a-d first present the normalized $x$-component of the electric field before any invisibility gene is injected. Here, a TE-polarized plane wave of frequency $f_0$ = 1.7 GHz travels along the $+z$-direction and impinges on dielectric objects with the same permittivity $\varepsilon_o$ = 6 whose shapes are the letters "A," a crescent, an umbrella, and the Chinese character "中," respectively. The simulated results in Fig. 3a-d show that all these objects scatter the incident wave obviously: the outgoing wavefronts are no longer planar, shadows appear at various positions, pronounced field disturbances occur near the object boundaries, and the global electric-field distribution deviates completely from that of a plane wave in free space.

Next, the invisibility-gene procedure of Fig. 1 is applied to cloak each object. For modeling convenience, every original shape in Fig. 3a-d is already discretized into subwavelength cubical units of side length $L$ = 0.15 $\lambda_0$. Following the previously described design method, we implant a spherical invisibility gene into each subwavelength unit and then reassemble the units into the same macroscopic shapes (A, crescent, umbrella, and "中"). The resulting modified objects are illuminated again by the same TE-polarized plane wave of the same frequency $f_0$ = 1.7 GHz from the same $+z$-direction. The corresponding normalized $x$-component of the electric field distributions are displayed in Fig. 3e-l. Panels 3e-h use the first strategy of structure-I with permittivity $\varepsilon_I$ = -1.88, whereas panels 3i-l adopt the second strategy of structure-I with permittivity $\varepsilon_I$ = 2.86.

Compared with the original objects without invisibility genes in Fig. 3a-d, both gene-injection schemes dramatically suppress scattering. In Fig. 3e-l the outgoing wavefronts exhibit only minimal distortion—essentially remaining planar and casting no discernible shadow—while extremely weak perturbations are confined to the immediate vicinity of the object contours. Consequently, the electric-field distribution across the entire computational domain almost perfectly recreates that of a plane wave in free space. Therefore, for both permittivity strategies, injecting the invisibility gene into identically shaped objects yields pronounced cloaking: the incident plane wave remains almost undisturbed, and scattering is effectively cancelled. Fig. 3 therefore numerically demonstrates that arbitrarily shaped dielectric objects can be cloaked by decomposing the target into subwavelength units, implanting a tailored invisibility gene into each unit, and then reassembling the units into the original geometry. Additionally, the simulation results demonstrate that using the invisibility gene injection method presented in Fig. 1, ideal invisibility can also be achieved for a TE-polarized plane electromagnetic wave of frequency $f_0$ incident from arbitrary directions (see **Supplementary Note 2** and **Movie 2**).

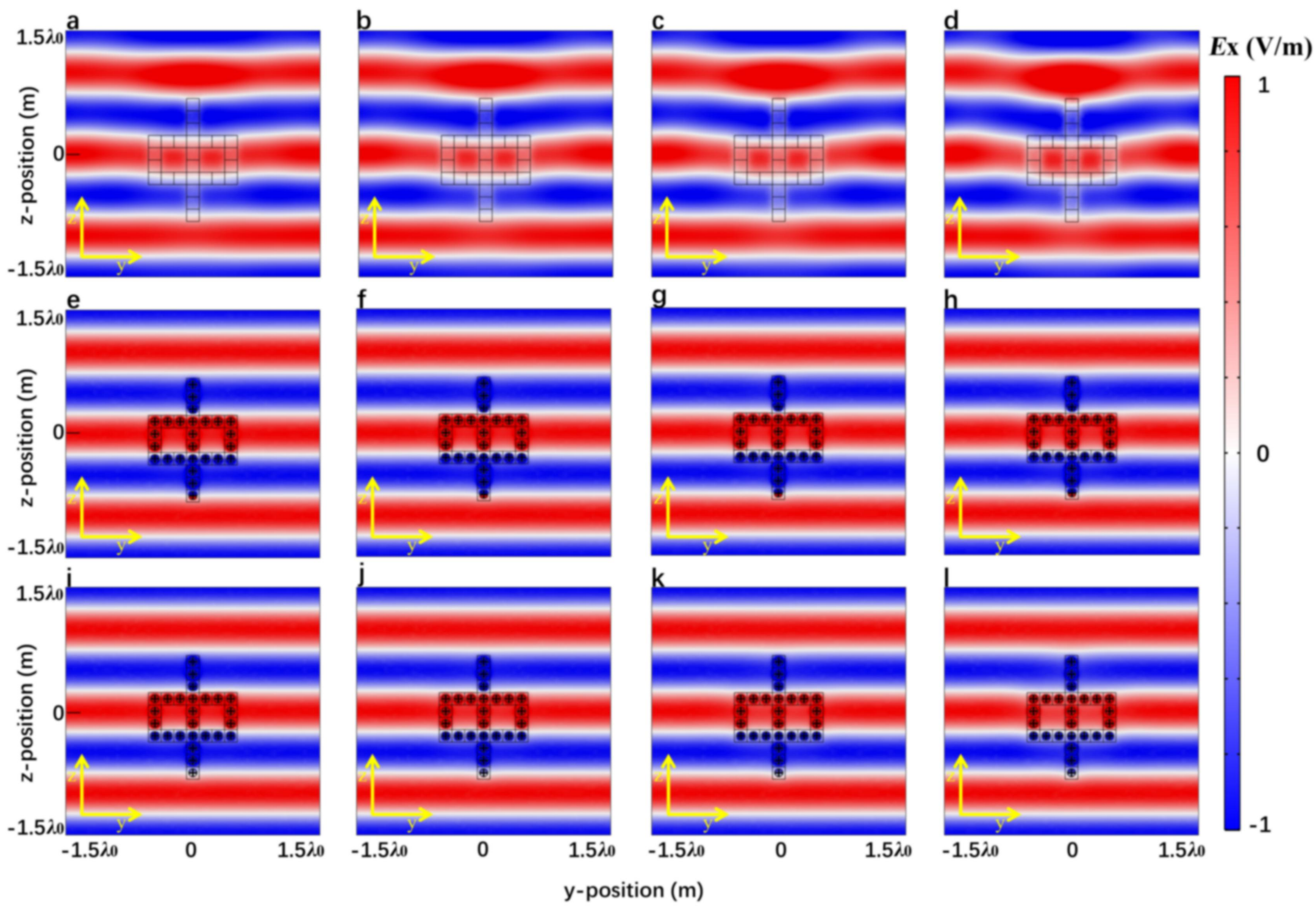

**Fig. 4| 3D simulated results: normalized simulated electric field's x-component $E_x$ on the x = 0 cross-section when a TE-polarized plane electromagnetic wave of the frequency $f_0$ = 1.7 GHz is incident along the positive *z*-axis onto structures with different material properties. a-d**, Structures shaped like the Chinese character "中", made of dielectrics with permittivity $\varepsilon_o = 3$, $\varepsilon_o = 4$, $\varepsilon_o = 5$, and $\varepsilon_o = 7$, respectively, without injected invisibility genes. **e-h**, Corresponding structures to **a-d** after injecting invisibility genes using the first strategy (Structure-I with negative permittivity in Fig. 2b). **i-l**, Corresponding structures to **a-d** after injecting invisibility genes using the second strategy (Structure-I with positive permittivity in Fig. 2b). In simulations, boundary conditions, geometric parameters, and material settings for the original objects and invisibility genes are detailed in the **Numerical Method**.

To further verify the universality of the proposed invisibility gene injection method for dielectric targets with different permittivities, we performed full-wave numerical simulation verification on targets with the same geometric shape but different permittivities, and the results are shown in Fig. 4. A TE-polarized plane electromagnetic wave with a frequency of $f_0$ = 1.7 GHz is incident along the +*z*-direction and irradiates four dielectric targets with the identical shape of the Chinese character "中" but permittivities of $\varepsilon_o = 3$, $\varepsilon_o = 4$, $\varepsilon_o = 5$, and $\varepsilon_o = 7$, respectively. When no invisibility gene is injected, the distribution of the normalized *x*-component $E_x$ of the electric field on the $x = 0$ cross-section is shown in Fig. 4a-d. The simulation results show that all targets produce a significant scattering effect on the incident electromagnetic wave: the outgoing wavefronts lose the characteristics of plane waves, obvious shadow regions appear behind the targets, strong field disturbances exist near the target boundaries, and the electric field distribution in the entire computational domain deviates significantly from that of a plane wave in free space.

Subsequently, we applied the invisibility gene injection procedure shown in Fig. 1 to design the cloaking for the above targets. For modeling convenience, all targets in Fig. 4a-d are discretized into subwavelength cubical units with a side length of $L = 0.15\lambda_0$. Following the design method proposed above, we implant a spherical invisibility gene matching its permittivity into each subwavelength unit, and then reassemble the units into the original macroscopic structure shaped like the Chinese character "中". Simulations are performed on the targets after invisibility gene injection under the same incident conditions, and the obtained distributions of the normalized x-component of the electric field are shown in Fig. 4e-l.

Among them, Fig. 4e-h adopt the first design strategy in Fig. 2b (Structure-I with negative permittivity): for the units with $\varepsilon_o$ = 3, 4, 5, and 7, the permittivity $\varepsilon_I$ of Structure-I is set to -0.89, -1.22, -1.55, and -2.23, respectively; Fig. 4i-l adopt the second design strategy (Structure-I with positive permittivity), with the corresponding $\varepsilon_I$ set to 3.88, 3.54, 3.20, and 2.51, respectively.

Compared with the original targets without invisibility genes in Fig. 4a-d, both gene injection schemes achieve an excellent scattering suppression effect. As shown in Fig. 4e-l, after the incident plane wave penetrates the targets, the outgoing wavefronts have only negligible distortion, essentially maintaining a perfect planar wave shape without obvious shadow regions; only extremely weak field perturbations are confined to the immediate vicinity of the target contours, and the electric field distribution in the entire computational domain is almost completely consistent with that of a plane wave in free space. The above results show that for dielectric targets with the same shape but different permittivities, excellent cloaking performance can be achieved only by injecting a customized invisibility gene according to the permittivity of each unit, the incident plane wave is almost undisturbed, and the scattering of the targets is effectively cancelled. Therefore, the numerical simulation results in Fig. 4 fully prove that the proposed method in this study can achieve efficient scattering-cancellation cloaking for dielectric targets with different permittivities by decomposing the targets into subwavelength units, injecting a customized invisibility gene matching the material properties of each unit, and then reconstructing the units into the original geometric structure, which verifies the universality of this method for targets with different material properties.

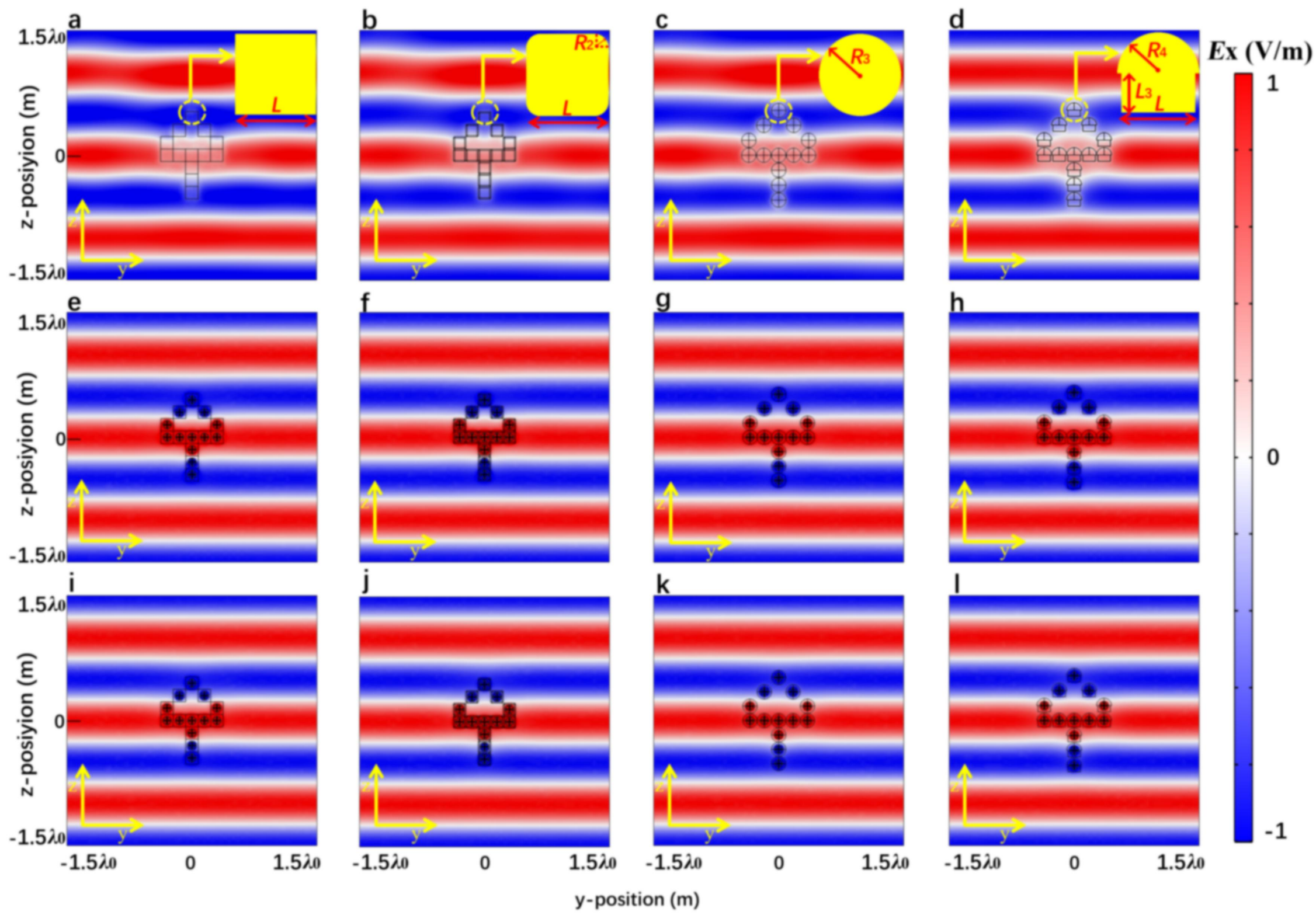

**Fig. 5 | 3D simulated results: normalized simulated electric field' s *x*-component $E_x$ on the $x = 0$ cross-section when a TE-polarized plane electromagnetic wave of the frequency $f_0 = 1.7$ GHz is incident along the positive *z*-axis onto structures composed of subwavelength units with different boundary shapes. a-d**, Arrow-shaped structures composed of subwavelength units with cubical, rounded cubical, spherical and mushroom-shaped boundaries, respectively. All structures are made of dielectric with permittivity $\varepsilon_o = 5$ without injected invisibility genes. **e-h**, Corresponding structures to **a-d** after injecting invisibility genes using the first strategy (Structure-I with negative permittivity in Fig. 2b). **i-l**, Corresponding structures to a-d after injecting invisibility genes using the second strategy (Structure-I with positive permittivity in Fig. 2b). In simulations, boundary conditions, geometric parameters, and material settings for the original objects and invisibility genes are detailed in the **Numerical Method**.

In practical application scenarios, the discretized subwavelength units of the target to be cloaked are not all regular cubical structures. To further verify the adaptability of the proposed invisibility gene injection method to subwavelength units with different boundary shapes, we performed full-wave numerical simulation verification on macroscopic targets composed of subwavelength units with different boundary shapes, and the results are shown in Fig. 5. A TE-polarized plane electromagnetic wave with a frequency of $f_0 = 1.7$ GHz is incident along the $+z$-direction and irradiates four dielectric targets with the identical arrow shape and the same permittivity $\varepsilon_o = 5$, which are composed of subwavelength units with four different boundary shapes: cubical, rounded cubical, spherical, and mushroom-shaped, respectively. When no invisibility gene is injected, the distribution of the normalized *x*-component $E_x$ of the electric field on the $x = 0$ cross-section is shown in Fig. 5a-d. The simulation results show that all targets produce a significant scattering effect on the incident electromagnetic wave: the outgoing wavefronts lose the characteristics of plane waves, strong field disturbances occur near the target boundaries, and the electric field distribution in the entire computational domain deviates significantly from that of a plane wave in free space.

Subsequently, we applied the invisibility gene injection procedure shown in Fig. 1 to design the cloaking for the above targets. Following the design method proposed above, we implant a spherical invisibility gene matching the material properties into each subwavelength unit with different boundary shapes, and then reassemble the units into the original arrow-shaped macroscopic structure. Simulations are performed on the targets after invisibility gene injection under the same incident conditions, and the obtained distributions of the normalized x-component of the electric field are shown in Fig. 5e-l. Among them, Fig. 5e-h adopt the first design strategy in Fig. 2b (Structure-I with negative permittivity): for the units with $\varepsilon_o = 5$, the permittivity $\varepsilon_I$ of Structure-I is uniformly set to -1.55; Fig. 5i-l adopt the second design strategy (Structure-I with positive permittivity), with the corresponding $\varepsilon_I$ uniformly set to 3.20. Compared with the original targets without invisibility genes in Fig. 5a-d, both gene injection schemes achieve an excellent scattering suppression effect. As shown in Fig. 5e-l, after the incident plane wave penetrates the targets, the outgoing wavefronts have only negligible distortion, essentially maintaining a perfect planar wave shape without obvious shadow regions; only extremely weak field perturbations are confined to the immediate vicinity of the target contours, and the electric field distribution in the entire computational domain is almost completely consistent with that of a plane wave in free space.

The above results show that even if the target to be cloaked is decomposed into subwavelength units with different boundary shapes, excellent cloaking performance can be achieved only by injecting a standardized and matched invisibility gene into each unit, the incident plane wave is almost undisturbed, and the scattering of the targets is effectively cancelled. Therefore, the numerical simulation results in Fig. 5 fully prove that the invisibility gene injection method proposed in this study has good universality for subwavelength units with different boundary morphologies, which further breaks through the strict constraints on the unit structure inherent in traditional methods, and provides reliable support for the cloaking design of targets with arbitrarily complex morphologies.

**Experimental design and results**

The previous numerical simulation results have confirmed that the core physical mechanism of the electromagnetic invisibility gene injection method proposed in this study for subwavelength units with different boundary shapes lies in the precise cancellation of electric dipole moments, which provides a theoretical basis for the structural optimization and engineering fabrication of invisibility genes. In the preliminary design, we all adopted spherical invisibility genes to achieve scattering cancellation. To meet the process requirements of subsequent experimental fabrication, we first analyzed the effect of corner radian passivation of spherical invisibility genes on the cloaking performance via numerical simulations (see **Supplementary Note 3** for details). The simulation results show that even if the spherical invisibility genes are optimized into cubical invisibility genes which are easier to fabricate, only minor adjustments to their internal geometric parameters are required to maintain ideal scattering cancellation and cloaking performance at the target operating frequency (see **Supplementary Note 4** for details), which fully verifies the structural flexibility and engineering practicability of the proposed method.

Based on the above conclusions, we conducted experimental verification in the microwave band to confirm the effectiveness of the electromagnetic invisibility gene injection cloaking method proposed in this study. In the experiment, to facilitate sample fabrication, we set the central operating frequency to $f_0 = 1.5$ GHz, corresponding to a free-space operating wavelength of $\lambda_0 = 20.0$ cm. The original target to be cloaked selected in this experiment (exhibiting an overall columnar asymmetric structure, as shown on the left side of Fig. 6a) is assembled from 21 standard subwavelength cubical units. Each unit has a side length of $L = 0.15\lambda_0$ and is fabricated from ceramic blocks with a permittivity of $\varepsilon_o = 6.8$ ($H_1$ = 9.0 cm, $W_1$ = 6.0 cm, $D_1$ = 9.0 cm). Among the cubical invisibility genes designed for the experimental samples, Structure-I is equivalent to an air medium at the operating frequency and requires no additional fabrication. Structure-II uses metallic zinc fabricated by direct cutting, and Structure-III uses aluminum alloy processed by 3D metal printing.We injected the fabricated cubical invisibility genes one by one into each original subwavelength unit of the target to be cloaked, and then reassembled them according to the original morphology. Finally, the complete sample with injected invisibility genes is obtained (as shown in the middle of Fig. 6a). The inset on the right side of Fig. 6a shows the disassembled structure of a single unit from the sample with injected invisibility genes, which clearly demonstrates that each unit is a standard cubical structural unit assembled from three parts: IV, V, and VI ($L$ = 3.0 cm, $L_1$ = 2.0 cm) (see **Supplementary Note 5** for details of sample fabrication).

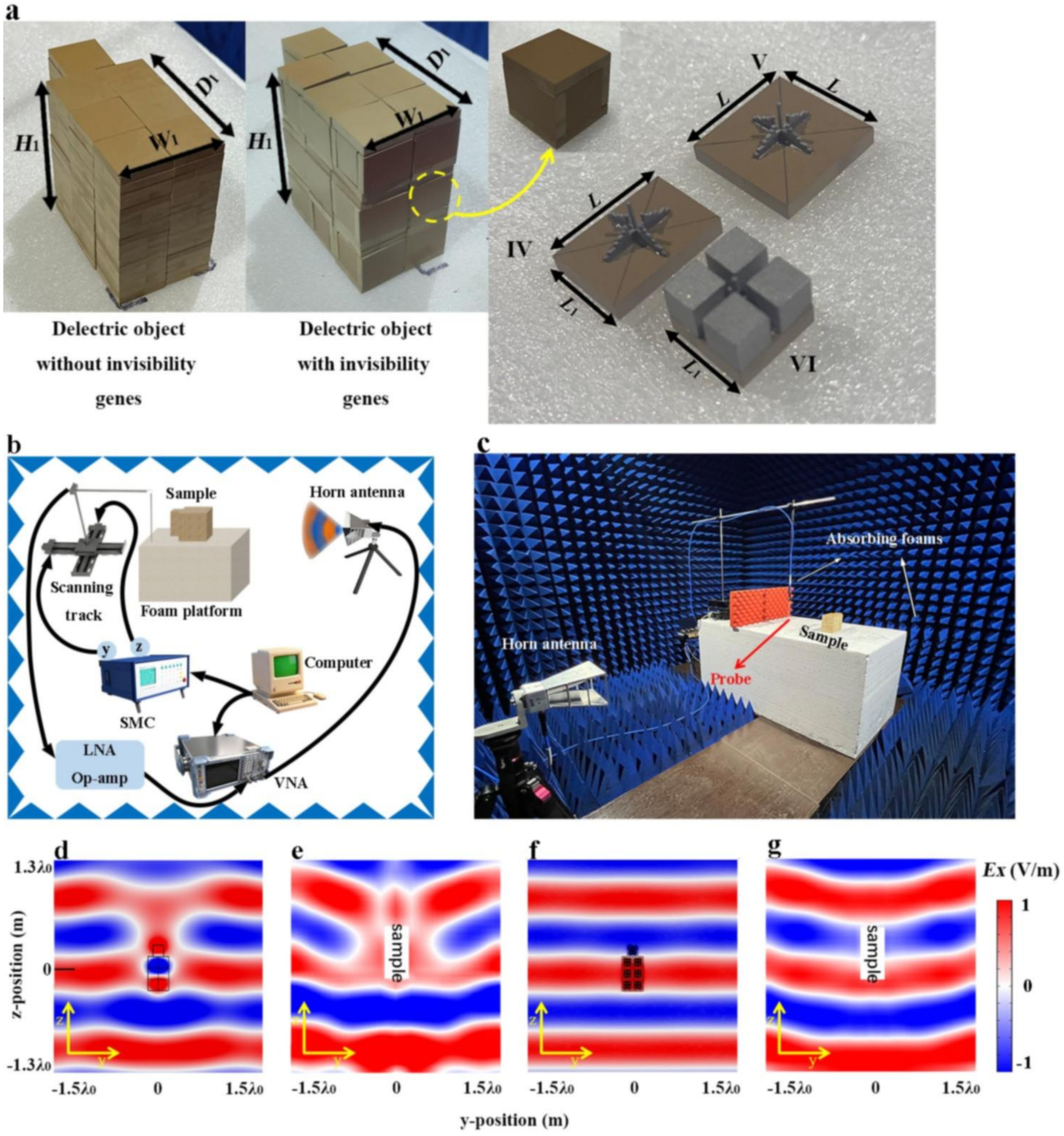

**Fig. 6| Experimental validation of cloaking via electromagnetic invisibility gene injection. a** Photographs of invisibility gene injection cloaking samples: (left) original reference sample, (middle) cloaked sample with customized cubical invisibility genes, (right inset) disassembled single unit composed of components IV, V, and VI, corresponding to air-equivalent Structure-I, 3D-printed zinc Structure-II and aluminum alloy Structure-III respectively. **b** Schematic of the microwave measurement setup. **c** Photograph of the experimental environment in the microwave anechoic chamber. **d** Simulated normalized electric field distribution of the $E_x$ component, under the incidence of a TE-polarized plane electromagnetic wave along the positive $z$-axis onto the object without invisibility gene injection. **e** Experimentally measured normalized electric field distribution of the $E_x$ component for the object without invisibility gene injection under the same incidence condition. **f** Simulated normalized electric field distribution of the $E_x$ component, under the incidence of a TE-polarized plane electromagnetic wave along the positive $z$-axis onto the object with invisibility gene injection. **g** Experimentally measured normalized electric field distribution of the $E_x$ component for the object with invisibility gene injection under the same incidence condition.

After the sample fabrication is completed, the microwave measurement system is constructed as schematically shown in Fig. 6b, and the photograph of the actual experimental setup is presented in Fig. 6c. The entire measurement setup is placed in a microwave anechoic chamber to eliminate the interference of environmental electromagnetic reflections on the measurement results. The sample is fixed on a low-permittivity foam platform with a height of 1.0 m to avoid the influence of ground reflections on electromagnetic wave propagation. An $x$-polarized horn antenna is used as the TE-polarized electromagnetic wave source, which is connected to port 1 of a vector network analyzer (VNA, ROHDE&SCHWARZ ZVL13). The central measurement frequency of the instrument is set to $f_0 = 1.5$ GHz, with a measurement frequency range of 1 GHz to 2 GHz. The horn antenna is mounted on a tripod, 150.0 cm away from the measurement plane, and aligned with the sample center along the $z$-axis to make the wavefront of the electromagnetic wave in the measurement area as close to a plane wave as possible. A metal probe obtained by stripping the cladding of a coaxial cable is used as the detection probe to measure the distribution of the $x$-component $E_x$ of the electric field in the 60.0 cm × 52.5 cm air region in the $y$-$z$ plane around the sample. The metal tip of the detection probe is placed along the $x$-direction to accurately detect the $E_x$ component, and the other end of the probe is connected to port 2 of the vector network analyzer via a coaxial cable. The detection probe is fixed on a suspended bracket (4.5 cm above the foam platform, at the mid-height of the sample), which is linked to a two-dimensional scanning track movable along the $y$-$z$ directions (see **Supplementary Note 6**).

During the scanning process, a self-written computer driver program controlled the stepper motor controller to drive the two-dimensional scanning track to move along the $y$-axis and $z$-axis with a step size of 1.5 cm to complete the electric field scanning measurement of the target area. The metal tip of the detection probe is always kept in the $x = 0$ plane (i.e., the mid-height plane of the sample), consistent with the simulation cross-sections in Figs. 3 to 5 above. Finally, the amplitude and phase information of the local electric field $E_x$ component are extracted from the S21 parameters measured by the vector network analyzer, and imported into the computer for subsequent data processing to obtain the electric field $E_x$ distribution map in the $x = 0$ plane (see **Supplementary Note 7** and **Supplementary Movie 3** for detailed experimental procedures).

The comparison of simulated and experimental results of the normalized electric field distribution of the sample before and after the injection of cubical invisibility genes is shown in Figs. 6d-g. Among them, Figs. 6d (simulated result) and 6e (experimental result) show the normalized electric field $E_x$ distribution of the control sample without invisibility gene injection. It can be clearly observed that after the incident electromagnetic wave passes through the sample, the outgoing wavefront is severely distorted, and an obvious scattering shadow area appears behind the sample, which is consistent with the scattering characteristics of the uncloaked target in the previous simulations. In contrast, Figs. 6f (simulated result) and 6g (experimental result) show that after injecting the customized cubical invisibility genes into the sample, the outgoing wavefront basically maintains a plane wave shape, the scattering effect is significantly suppressed, and the electric field distribution in the entire measurement area is highly consistent with the propagation characteristics of a plane wave in free space, confirming the effective cancellation effect of the invisibility genes on the target scattering. The experimental measurement results are in good agreement with the numerical simulation results. The minor differences between them mainly stem from the limitations of experimental conditions: the limited distance between the horn antenna and the sample results in a non-ideal plane wave incident wavefront, and there are non-ideal factors such as fabrication tolerances and environmental clutter in the experiment. Overall, the microwave experimental results fully confirm the feasibility and effectiveness of the electromagnetic invisibility gene injection method proposed in this study in practical scenarios.

**Conclusion**

This work establishes a universal unit-level electromagnetic invisibility gene injection scheme that realizes efficient scattering-cancellation cloaking for arbitrarily shaped and large-scale dielectric objects. The invisibility gene module achieves complete scattering cancellation by synergistically regulating the electromagnetic response of each subwavelength unit, and only requires the adjustment of key dielectric parameters to adapt to different host materials. Full-wave simulations consistently confirm excellent cloaking performance for arbitrary macroscopic shapes, a wide range of material permittivities, and diverse subwavelength unit morphologies. Microwave-band experimental results clearly demonstrate significant scattering suppression and effective electromagnetic invisibility claoking effect, which fully validate the practical effectiveness of the proposed method. This approach preserves the electromagnetic perception capability of the cloaked object, breaks the strict limitations of traditional scattering-cancellation cloaking methods on shape, size, and material adaptability. Numerous practical structures are assembled from basic dielectric units, such as buildings stacked by dielectric blocks, radar supports, and communication base station pillars. By integrating invisibility genes during manufacturing, such structures maintain their original mechanical and supporting functions while achieving electromagnetic invisibility. This technology exhibits broad application potential in radar systems, 5G base stations, and other electromagnetic communication scenarios. It can also be extended to electromagnetic transparent protective covers to shield equipment from wind, sand, rain and corrosion without interfering with electromagnetic signal transmission, thereby improving the stability and service life of electromagnetic equipment in practical environments.

## Methods

### Numerical Method.

To verify the performance of the invisibility genes for cloaking different dielectric objects of arbitrary shapes, all numerical simulations in this study are conducted using COMSOL Multiphysics 5.6 with license number 9406999. All simulations are 3D cases, where the wave optics module with the steady-state solver is selected to simulate the electromagnetic wave scattering from objects with or without injected invisibility genes. A free tetrahedral meshing is used, with a maximum grid size of $\lambda_0/10$. The working frequency is set to $f_0 = 1.7$ GHz, corresponding to a wavelength of $\lambda_0 = 0.176$ m. For the simulations in Figs. 3 to 5, all outer boundaries of the computational domain are set as scattering boundary conditions to simulate an infinite free-space environment; the background incident field is set as a unit-amplitude TE-polarized plane electromagnetic wave, with the electric field polarized along the $x$-direction and propagating along the positive $z$-axis.

For the cubical subwavelength units obtained by discretizing the target (Figs. 3 and 4), each unit is a non-magnetic dielectric cube with a side length of $L = 0.15\lambda_0$ and a permittivity of $\varepsilon_o$. For the subwavelength units with irregular boundaries in Fig. 5, the geometric parameters are set as follows: the rounded cubical unit has a fixed side length of $L$ with a fillet radius of $R_2 = L/10$; the spherical unit has a fixed radius of $R_3 = \sqrt{3}\,L/3$; the mushroom-shaped unit has a hemispherical head with a radius of $R_4 = 3L/5$ and a rectangular shank with dimensions of $L\times L\times L_3$ ($L_3 = L/2$). Regardless of whether the subwavelength unit is cubical, rounded cubical, spherical, or mushroom-shaped, the geometric parameters of the injected spherical invisibility gene are uniformly set as: $R = L/3$, $R_1 = 71L/240$, $H = L/2$, $W = L/40$, $P = 3L/40$, $D = 2R = 2L/3$, $D_1 = \sqrt{7}\,L/3$. The material parameters of the spherical invisibility gene are set as follows: Structure-II and Structure-III are metallic materials, which are set as perfect electric conductors (PEC) in microwave band simulations; Structure-I is a dielectric with tunable permittivity $\varepsilon_I$, whose value is determined by the permittivity $\varepsilon_o$ of the host subwavelength unit (see Fig. 2b and **Supplementary Note 1** for the mapping relationship between $\varepsilon_o$ and $\varepsilon_I$).

## Acknowledgements

National Natural Science Foundation of China (No. 12274317, No. 12374277), San Jin Talent Support Program—Shanxi Provincial Youth Top-notch Talent Project, the Natural Science Foundation of Shanxi Province (202303021211054), Shanxi Province Higher Education Institutions Young Faculty Research and Innovation Support Program (2025Q006), and Shanxi Province Postgraduate Practical Innovation Project.